\documentclass[a4paper,10pt]{article}
\usepackage[utf8]{inputenc}
\usepackage{amssymb}
\usepackage{amsfonts}
\usepackage{amsmath}
\usepackage{cite}
\usepackage{doi}
\usepackage{hyperref}

\newcommand{\vev}[1]{ \left\langle #1 \right\rangle}
\newcommand{\Ukin}{U_{\mathrm{kin}}}
\DeclareMathOperator{\re}{Re}

\begin{document}

\title{A Tree Swaying in a Turbulent Wind: A Scaling Analysis}
\author{Theo Odijk \\
{\small Lorentz Institute for Theoretical Physics}\\
{\small University of Leiden
Niels Bohrweg 2
2333 CA Leiden, The Netherlands}\\
{\small e-mail: odijktcf@online.nl}}
\maketitle

\begin{abstract}
A tentative scaling theory is presented of a tree swaying in a turbulent
wind. It is argued that the turbulence of the air within the crown is in the
inertial regime. An eddy causes a dynamic bending response of the branches
according to a time criterion. The resulting expression for the penetration
depth of the wind yields an exponent which appears to be consistent with that
pertaining to the morphology of the tree branches. An energy criterion shows
that the dynamics of the branches is basically passive. The possibility of
hydrodynamic screening by the leaves is discussed.
\end{abstract}

\section{Introduction}
The time criterion introduced by Lumley~\cite{1,2,3} plays a central role in the
theory of drag reduction in turbulent flow. Drag reduction implies a
reduction in the friction factor in pipe flow experiments. The time scale of
turbulence at some spatial scale is compared with the main reaction time of
a polymer chain, for instance. When they are equal, this signals a change in
the dynamical behavior of the solution because the chain becomes markedly
deformed. A turbulent flow acting on a chain is a bit similar to the action
of a slightly damped spring. Nevertheless, the two systems may not be viewed as
identical: the turbulent time scale is merely a statistical correlation time
$\tau_{c}$. Beyond $\tau_c$ the fluid is effectively decorrelated thus the chain is assumed
to be swept along with the flow in some random direction once every period $\tau_c$.
The ratio of the two scales also figures in more recent
elaborate theoretical work on drag reduction~\cite{4,5,6}.

A major difficulty in understanding drag reduction has been that it is a
phenomenon whose relative magnitude is never more than order unity. It is
therefore of interest to study systems of a complexity close in spirit to
that of polymers but where the object-flow interaction is more amenable to
analysis by experiment. A tree swaying under the influence of a turbulent
wind is a case in point. Here, I apply a time criterion to this problem
inspired by de Gennes' discussion of a polymer chain deformed by bulk
turbulence~\cite{4, 5}.

The deformation of trees modeled as elastic structures under winds has been
investigated extensively recently~\cite{7,8,9,10,11,12,13,14}. However, it is difficult to
simulate wind turbulence at extremely high Reynolds numbers. Accordingly, it
is useful to present a scaling analysis of a tree swaying in a turbulent
wind to try to delineate the relevant dimensional parameters involved.

\section{Time Criterion}
Let us consider a tree of height $H$ swaying under the influence of a moderate
wind of average speed $U$. Simple observations by the author in the field
prove that its branches oscillate rather haphazardly and quite independently
of each other. The aerodynamical flow circulating within the crown may be
argued to be turbulent. The Reynolds number $\re_{0}=HU/\nu$ of a tree
of height $H\simeq 10$m, say, swaying in a wind of speed $U\simeq 10$m/s is
a formidable $10^{7}$ (the kinematic viscosity $\nu $ of air is about $%
0.15$cm$^2$/s). The Kolmogorov dissipation scale $\lambda =H/\re^{3/4}$%
\cite{15} is then only a minute $0.1$mm. I assume the turbulent cascade is
essentially unperturbed by the tree branches and leaves (see below). The
tree is thus termed a passive object. Yet the Reynolds number $\re_{d}=dV_{d}/\nu$ associated with some branch of diameter d equal to about
$1$cm, say, is still not too large for Karman vortices to separate from its
surface ($\re_d\simeq 10^{2}$ ; see ref.~\cite{16}). The velocity of the wind
is split up into a systematic component and a fluctuational component $%
\vec{V}$. There is a stick boundary condition at ground level but
the systematic profile is not the usual logarithmic function of height if
the tree were absent~\cite{17} (For a forest of trees, a phenomenological,
so-called exponential profile is known to hold~\cite{17} because the canopy does
exert some influence on the wind profile). For winds blowing on a flat
landscape, the fluctuational term $\vec{V}(\vec{r})$
is known to be isotropic on empirical grounds as first shown by G. I.
Taylor [18]. There is no reason to suspect this would be otherwise if the
average profile were different. The fluctuational speed $V_{d}$ of the
air near a branch is about $U( d/H) ^{1/3}$.

In view of the nonalignment of the branches of a tree, the total sum of
Karman vortices arising from the interaction of the wind with the branches
may be supposed to be a random variable. On the whole, one concludes that
the air circulating throughout the crown is in the inertial regime. On
dimensional grounds, the fluctuational velocity obeys Kolmogorov's relations
like
\begin{equation}
V( r) \simeq ( \varepsilon r) ^{1/3},  \label{eq1}
\end{equation}
where the magnitude $V( H) =\mathcal{O}( U) $ and the rate of
dissipation $\varepsilon $ is of order $U^{3}/H$. In the inertial regime
pertaining to scales much larger than the dissipation scale $\lambda _{1}$
it is assumed dissipation is absent~\cite{15}. Equation~\eqref{eq1} follows from
dimensional analysis.

I anticipate that the turbulent wind penetrates dynamically into an outer
shell-like region of the tree crown of thickness $\ell _{\ast }$ (which
will be termed the penetration depth). Thus, branches within this region are
excited but those in the inner core are quiescent. I note that in practice
the length $\ell( d) $ of a branch measured from the tip increases
monotonically with $d$. The tip itself must have a nonzero diameter $d_{0}$
otherwise the branch could not bear a bud at its end. The bud forms new
leaves, higher order branches and causes the main branch to be elongated
each year. Accordingly, a Taylor expansion of $d( \ell) $ ought to
be possible at small $\ell$%
\begin{equation}
d(\ell) =d_0\left( 1+\alpha_1\ell+\alpha_2\ell^{2}+\dots \right)
\label{eq2}
\end{equation}%
with empirical coefficients $\alpha_1$, $\alpha_2$, etc. By contrast,
at large $\ell$ one expects a fractal structure~\cite{19}%
\begin{equation}
d\sim \ell^{\beta }  \label{eq3}
\end{equation}%
An eddy of the turbulent air encompassing a section of branch of length $\ell$
typically has a characteristic velocity $V(\ell) \simeq (
\varepsilon \ell) ^{1/3}$ ( eq.~\eqref{eq1}).
The time scale of the flow is%
\begin{equation}
\tau_{\ell}\simeq \frac{l}{V( \ell) }\simeq \frac{\ell^{2/3}}{\varepsilon
^{1/3}}\simeq \frac{H}{U}\left( \frac{\ell}{H}\right) ^{2/3}  \label{eq4}
\end{equation}%
In general, the size of the crown is of the order of the height of the tree~\cite{10}.

For the moment, let us suppose that a section of branch within the annular
region has a uniform diameter $d$. Then a time scale associated with the
bending oscillations of the branch is given by%
\begin{equation}
\tau_{b}\simeq \frac{\ell^{2}}{d}\left( \frac{\rho_w}{Y}\right)
^{1/2}\simeq \frac{\ell^{2}}{dS}  \label{eq5}
\end{equation}%

In effect the bending energy of the branch is $E\ell/2R_c^2$~\cite{20}
where $E$ is the bending modulus and $R_c$ is the radius of
curvature of a bent branch. The deflection $z$ of the branch from a straight
form is $z\simeq \ell^2/R_c$ and Hooke's modulus equals $k\simeq
E/\ell^3\simeq d^4Y/\ell^3$ where $Y$ is Young's modulus (here, for
simplicity within a scaling analysis, we consider the wood to be an
isotropic material with one characteristic modulus). The section of branch
has a mass $m\simeq d^{2}\ell\rho_w$ with $\rho_w$ the density of the
wood and the characteristic time scale of a bending oscillation is $\tau_b=(m/k) ^{1/2}$ yielding eq.~\eqref{eq5}. The
speed of sound in wood is $S=\left( Y/\rho_w\right) ^{1/2}$.

Ultimately, at some penetration depth $\ell_{\ast }$, the bending oscillation
of the branch is in concert with turbulent eddies acting upon it. Accepting
the time criterion $\tau_{\ell}\simeq \tau_d$, we have%
\begin{equation}
d_{\ast }\simeq \frac{\ell_{\ast }^{4/3}}{H^{1/3}}\left( \frac{U}{S}\right)
\label{eq6}
\end{equation}%

This is interpreted as follows:
\begin{enumerate}
 \item At large $\ell$, we require $\beta >4/3$ if the argumentation is to be
self-consistent. Hence, the scaling analysis predicts a lower bound on the
exponent $\beta $. Experimental estimates for $\beta $ have been quoted~\cite{10}:
$\beta =1.37$ for pine trees and $\beta =1.38$ for walnut trees. This
is not strictly true for the tree structures are regarded as discrete
fractals whereas eq.~\eqref{eq3} is continuous. These
experimental exponents are surprisingly close to the theoretical minimum
value.
\item At low $\ell$, we need to know the value of the coefficients $\alpha_1$
and $\alpha_2$ to proceed. eqs.~\eqref{eq2} and \eqref{eq6} would indicate that there may sometimes be two
solutions which would conflict with the premise of the theory outlined here.
\item The typical aspect ratio of a branch predicted from eq.~\eqref{eq6}is of the order of $U/S$ which is consistent with values in
the field ($U=10$m/s; $S= 3000$ m/s).
\item Eq.~\eqref{eq6} is only correct if it is very slowly
varying ($\alpha_1 d_0\ll 1$). A detailed analysis bears this out (using the
results of ref.~\cite{21}).
\item There are of course smaller branches attached to a given branch and
so forth but their mass is relatively small. Equation~\eqref{eq6}
remains valid to the leading order.
\item The mass of the leaves attached is also of the order of the mass of
a branch~\cite{11, 12}. This implies eq.~\eqref{eq6} is correct to
within $\mathcal{O}( 1) $.
\end{enumerate}

A cluster of leaves on a branch may be viewed as a porous medium so
one expects the flow surrounding a test leaf to be screened somehow by the
blocking of the aerodynamics by other leaves. At intermediate Reynolds
numbers, a quadratic nonlinearity may be retained in the Navier-Stokes
equation to set up a convenient nonlinear version of the Brinkman screening
theory~\cite{22}. Recently, however, a modification of the Oseen approximation~\cite{23} has been proposed to increase its range of validity well beyond $\re=\mathcal{O}(1)$~\cite{24}. The ruse is to use a higher renormalized viscosity (here denoted by $\nu$). Thus, the velocity satisfies the modified Oseen equation, this will be used in section~\ref{sec:4}.

In the next section I discuss the possible impact of the elastic tree
energy on the aerodynamic turbulence.

\section{Energy Criterion}
In the analysis above, the tree has been regarded as a passive entity. In
order to asses whether this assumption is correct, I next consider an energy
criterion as previously introduced by de Gennes in the case of drag
reduction by polymer~\cite{4, 5}. A turbulent eddy of size $\ell$ and volume $\ell
^3$ is assumed to fluctuate coherently and isotropically on a scale $\ell$
(it also encompasses smaller eddies fluctuating at shorter time scales).
Hence, the kinetic energy density (equal to the magnitude of the Reynolds
stress) of such an eddy is $\rho_{a}V^{2}( \ell)$ where $\rho_a$
is the density of air. It is rewritten with the help of eq.~\eqref{eq1}
\begin{equation}
\Ukin(\ell) \simeq \rho _{a}U^{2}\left( \frac{\ell}{H}\right) ^{2/3}
\label{eq7}
\end{equation}%

Now the bending energy of a branch is the average of $Ez^{2}/\ell^{3}$
and since $z$ is at most $\mathcal{O}(l) $, an upper bound on the bending
energy density is%
\begin{equation}
U_{b,\max }( \ell) \simeq \frac{d^{4}Y}{\ell ^{3}}  \label{eq8}
\end{equation}%

The section of the branch is enclosed by a blob of air of volume $%
\ell^{3}$. The ratio of the two densities is given by%
\begin{equation}
R=\frac{U_{b,\max }}{\Ukin}=\frac{\rho_w}{\rho_a}\left( \frac{d}{\ell}%
\right)^{2}  \label{eq9}
\end{equation}%
with the use of eq.~\eqref{eq6}. This is about $10^{-1}$ so
the branch is inferred to fluctuate passively. Nevertheless, a densely
branched tree may conceivably have $R=\mathcal{O}( l) $ if there are enough
branches within $\ell^{3}$.

\section{Leaf Aerodynamics}\label{sec:4}

The Reynolds number of air flow near a leaf in the tree is typically about a
hundred. Thus, we try to study the aerodynamics in the Oseen approximation~\cite{22}%
\begin{equation}
\frac{\partial \vec{V}}{\partial t}+\vec{U}\cdot
\vec{\nabla}\vec{V}=-\rho_{a}^{-1}\vec{\nabla}p+\nu \Delta \vec{V}  \label{eq10}
\end{equation}%
where $\vec{U}$ is the background velocity and $p$ is the
pressure. The air may be regarded as incompressible because $U$ is much
smaller then the velocity of sound in air. Let us set $\re\equiv 0$
momentarily and assume the cluster of leaves surrounding a branch gives rise
to a hydrodynamic screening length $\zeta $. A full analysis involving all
particles is complicated~\cite{25} but a compact treatment introducing $\zeta $
at the beginning and a Schwinger variational principle yields identical
results fast~\cite{26}. Here, a simple scaling analysis is given to see if
hydrodynamic screening between the leaves may exist. In the stationary
limit, we have upon enforcing screening via a Darcy term%
\begin{equation}
\nu \Delta \vec{V}-\frac{\nu }{\zeta ^{2}}\vec{V}=\rho_{a}^{-1}\vec{\nabla }p  \label{eq11}
\end{equation}%
The (pre-averaged) velocity perturbation by a point-like force $f$ (a delta
function) is then~\cite{25}%
\begin{equation}
V\sim \frac{e^{-r/\zeta }f}{\nu \rho_{a}r}  \label{eq12}
\end{equation}%
which is simply a screened form of the usual hydrodynamic decay. The
friction coefficient of a leaf, $\omega $ is then an average in terms of
the pair correlation function $g(\vec{r}) $~\cite{27}%
\begin{equation}
\omega ^{-1}\simeq \vev{\frac{e^{-1/\zeta }}{\nu \rho_a r}} \label{eq13}
\end{equation}%
For leaves viewed as platelets of surface area $S_p$, $g ( \vec{r}) $ scales as $r^{-1}$ so that the friction coefficient becomes%
\begin{equation}
\omega \simeq \frac{\nu \rho_{a}S_p}{\zeta }  \label{eq14}
\end{equation}%
If the swarm of leaves is enclosed in a column of length $l$ and
cross-section $A$ (volume $\Omega =lA$), the pressure difference $\Delta p$ is expressed by eq.~\eqref{eq11} (Darcy's law)%
\begin{equation}
A\Delta p=\Omega \nu \rho_{a}U/\zeta^{2}  \label{eq15}
\end{equation}%
On the other hand, this force on $N$ leaves is also given by eq.~\eqref{eq14}
\begin{equation}
A\Delta p=N\nu \rho_a S_p\zeta^{-1}U  \label{eq16}
\end{equation}%
We finally obtain an expression for the hydrodynamic screening length%
\begin{equation}
\zeta \simeq \frac{\Omega }{NS}=S^{1/2}\varphi ^{-1}  \label{eq17}
\end{equation}%
The variable $\varphi $ is a "hydrodynamic" volume fraction $S_p^{3/2}N/\Omega
$.

We next scale relevant terms in eqs.~\eqref{eq10} and~\eqref{eq11} by introducing an intermediate scale $r$ with swarm
size $\gg r\gg\zeta $. The impact of inertia is denoted by the dimensionless
quantity $J$%
\begin{equation}
J\equiv \frac{\text{inertial term}}{\text{screening term}}=\frac{U\zeta^{2}}{r\nu }  \label{eq18},
\end{equation}
though the Reynolds number pertaining to a single leaf in the flow field is%
\begin{equation}
\re_{\ell}\equiv \frac{US_p^{1/2}}{\nu }.  \label{eq19}
\end{equation}%
We therefore have%
\begin{equation}
\frac{J}{\re_{\ell}}=\frac{S_p^{1/2}}{r\varphi^2}  \label{eq20}
\end{equation}%
On expects $\varphi =\mathcal{O}( 1) $ or possibly even larger. Hence, the
ratio $J$ may be smaller than unity at substantial scales $r$. Inertia could
be neglected in that case so the computation of the screening length is
self-consistent. The flow at scale is laminarized by the strong screening
collectively caused by the leaves.

\section{Concluding Remarks}

The main expression derived here is eq.~\eqref{eq6} which
gives the penetration depth $\ell_{\ast }$ in terms of the branch diameter
$d_{\ast}$. The morphology of the tree imposes a second relation so that a
unique $\ell_{\ast }$ is obtained provided the exponent $\beta $ is not
lower than $4/3$. This is in accord with recent measurements of $\beta $
which are slightly above this number. Note that there are other theories
providing another estimate for $\beta $. Reasoning based on hydraulic
networks~\cite{28, 29} or an elasticity theory~\cite{19} yield $\beta =3/2$. On the
other hand, the fractal approximation deduced here is bound to break down at
small $\ell $. As mentioned earlier, a Taylor expansion like eq.~\eqref{eq2}
could lead to inconsistencies. However, we know that wood hardens because it
loses moisture over time~\cite{30}. The $\ell $ dependence of the time scale given by %
eq.~\eqref{eq5} would lessen if this effect were taken into account. Multiple
solutions could then be obviated.

The sealing picture presented here is only a zeroth-order theory. In
reality, the initial regime may be perturbed by dissipation arising from air
flow about the leaves. Canopies have been investigated in some detail. It is
found that Kolmogorov's laws do not always hold at large distances~\cite{31} with
implications for the restricted validity of eq.~\eqref{eq6}. Viscous damping~\cite{32}
within the wood of tree branches has also been neglected. Moreover, at
high wind speeds leaves may reconfigure and curl up ~\cite{33}.

\bibliographystyle{unsrturl}
\bibliography{Tree}
\end{document}